# Impacts of Climate Change on Photovoltaic Potential in Africa

Eva Lu[1], Dongdong Wang[2*]

1 International Department of Beijing No.80 High School, Beijing 100102, China

2 School of Earth and Space Sciences, Peking University, Beijing 100871, China

*Correspondence : Dongdong Wang

Address: No.5 Yiheyuan Road, Haidian District, Beijing 100871, P.R.China

**Abstract**

Africa holds the world's highest solar irradiance yet has <2% of global photovoltaic (PV) capacity, leaving 600 million people without electricity access. However, climate change impacts on its photovoltaic potential remain understudied. Using four decades of ERA5 reanalysis data (1980–2020) at 0.25° spatial resolution, we quantify the contributions of key climate factors to historical changes in African PV potential through multivariate decomposition. Continental PV potential increased by 3.2%, driven primarily by enhanced solar radiation (+0.52 W/m², accounting for $72\pm8\%$ of the gain), and partially offset by warming-induced losses (+1.2 °C, contributing $-23\pm5\%$). East Africa gained >6% from radiation enhancement, while North Africa declined by 0.5% as extreme heat (+2 °C) overwhelmed radiation benefits. Critically, stability analysis using the coefficient of variation (CV) reveals that high-irradiance subtropical zones are highly variable (CV=0.4), in contrast to stable equatorial regions (CV=0.1), challenging the assumption that resource abundance ensures reliability. These findings reframe Africa's solar strategy: North Africa requires prioritizing heat-resilient technology over capacity maximization; subtropical zones demand grid-storage co-investment; and East Africa presents globally competitive opportunities for rapid, stable deployment. By resolving spatiotemporal heterogeneities and quantifying climate-driver contributions, our analysis provides an actionable framework for climate-resilient solar deployment, critical for Africa's energy transition and climate mitigation.

## 1. Introduction

Achieving deep decarbonization of the global energy system requires a rapid transformation of the power sector through large-scale integration of renewable sources. Among these, photovoltaic (PV) technology has emerged as a cornerstone of the clean energy transition due to its scalability, modular design, and continuously declining cost (Zhang et al., 2022). However, long-term PV system performance remains critically dependent on climatic variables of surface-downwelling shortwave radiation (RSDS), air temperature (TAS), and wind speed (WS), all of which are undergoing significant change under global warming (Liu et al., 2025; Bamisile et al., 2025). Understanding how these climate variables influence PV output is essential for ensuring energy security, optimizing system design, and planning resilient energy transitions (Adenle, 2020; Obahoundje et al., 2024).

Africa represents a unique and urgent frontier in this context. The continent receives the highest levels of solar irradiance on Earth, offering a theoretical technical potential exceeding 10 TW (Neher et al., 2020; Sawadogo et al., 2020). However, Africa currently contributes less than 2% of global PV capacity, and hundreds of millions of people still lack access to modern electricity services (Ebhota & Tabakov, 2023). The anticipated surge in energy demand driven by population growth, industrialization, and urbanization underscores the urgency of accelerating solar deployment across the continent (Adenle, 2020; Yıldıran et al., 2025). At the same time, Africa is warming faster than the global average (Umeh & Gil-Alana, 2024), with energy infrastructure highly exposed to extreme heat, drought, and shifting precipitation patterns (Bichet et al., 2019; Bloomfield et al., 2022). These climatic stresses threaten to alter both the magnitude and stability of solar energy resources, posing a serious challenge to the reliability of future PV systems (Feron et al., 2020; Hou et al., 2021).

A growing body of research has investigated the influence of climate change on solar energy potential across different parts of the world. Crook et al. (2011) provided one of the earliest analyses of the combined effects of changing solar radiation and temperature on PV performance under global climate scenarios. Their study revealed that rising temperatures could reduce cell efficiency, partially offsetting potential gains from enhanced irradiance. Jerez et al. (2015) extended this work by examining Europe specifically, demonstrating significant temperature-induced losses in southern Europe despite abundant solar radiation. They highlight the importance of including multiple climatic variables rather than considering irradiance alone. Hasan et al. (2015) explored technological responses to thermal losses, showing that integrating phase-change materials could effectively mitigate overheating and sustain PV efficiency under elevated temperatures. Chen et al. (2024) conducted a global assessment using high-emission climate scenarios and concluded that while solar radiation may increase in many areas, these gains are frequently outweighed by efficiency losses from rising surface air temperatures, leading to complex and spatially heterogeneous patterns of PV productivity. Similar findings have been reported by Agbor et al. (2023) and Hudișteanu et al. (2024), emphasizing the sensitivity of PV efficiency to temperature and the importance of regional climate feedbacks. Collectively, these studies underscore that future PV output is not governed by a single climatic driver but results from the combined influences of multiple variables (Shaker et al., 2024; Sharaf et al., 2022).

Despite the above advances made at the global and regional scales, Africa remains markedly underrepresented in climate-solar research. Only limited studies have explicitly examined the continent's solar resource under changing climate conditions. Feron et al. (2020) analyzed the effects of climate extremes on PV generation at selected locations in Africa and identified significant production variations during heatwaves and dry spells. However, their analysis did not decompose the individual contributions of key variables such as radiation, temperature, and wind speed, which limits the understanding of underlying causes. Mokhtara et al. (2021) investigated PV performance in the arid regions of North Africa and found that high ambient temperatures significantly reduce conversion efficiency. However, their study focused on specific desert sites and lacked continental coverage, thereby constraining its applicability to Africa's diverse climatic zones. Other assessments have relied on coarse-resolution global climate models that are inadequate for representing the spatial gradients characteristic of the African environment (Bichet et al., 2019; Bloomfield et al., 2022). As a result, localized variations that strongly influence PV yield are frequently overlooked.

Filling this knowledge gap is particularly important because Africa's diverse climate produces sharp contrasts in PV performance. Equatorial regions experience consistently high humidity and dense cloud cover, suppressesing irradiance while maintaining moderate temperatures (Alemu et al., 2024). By contrast, subtropical regions have abundant sunshine, but also experience strong interannual variability and occasional heat extremes that degrade PV efficiency (Sawadogo et al., 2020; Long et al., 2024). Arid zones face additional challenges from dust storms and extreme thermal conditions, which can attenuate incoming radiation and accelerate module aging (Narváez et al., 2022). Understanding how these distinct climatic processes have historically shaped PV potential is vital for guiding both near-term investment and long-term resilience planning (Obahoundje et al., 2024; Yıldıran et al., 2025).

The present study provides a continent-wide quantitative assessment of how historical changes in key climate variables have influenced PV potential across Africa. Using four decades of ERA5 reanalysis data from 1980 to 2020 at 0.25° resolution (Hersbach et al., 2023), we employ a multivariate decomposition framework to quantify the relative contributions of RSDS, TAS, and WS to observed changes in PV potential. In addition to evaluating mean trends, we assess the temporal stability of solar resources through a coefficient-of-variation (CV) analysis. This integrative approach moves beyond the limitations of single-variable studies to provide a comprehensive view of the climatic mechanisms that shape Africa's solar energy potential. By resolving spatial heterogeneity and identifying dominant regional factors, our analysis establishes an evidence-based foundation for designing climate-resilient solar deployment strategies that account for both resource abundance and operational stability across the continent.

## 2. Data and Methods

### 2.1 Research Framework

Figure 1 presents the overall analytical framework of this study, which integrates climate data processing, PV potential modeling, driver decomposition, and stability assessment. The workflow consists of four interconnected stages. First, we compile four decades (1980–2020) of hourly climate data from ERA5 reanalysis at 0.25° resolution, focusing on three primary surface variablesof RSDS, TAS, and WS. Second, we calculate monthly PV potential using an empirically-derived model that accounts for radiation-temperature-wind interactions (Section 2.2). Third, we employ multivariate decomposition to isolate and quantify the individual contributions of RSDS, TAS, and WS to observed changes in PV potential (Section 2.3). Finally, we assess temporal stability through CV analysis and evaluate the impacts of extreme climate events including heatwaves, droughts, and dust storms (Sections 2.4–2.5). This framework enables both mechanistic understanding of historical climate impacts and strategic assessment of resource reliability across diverse African climate zones.

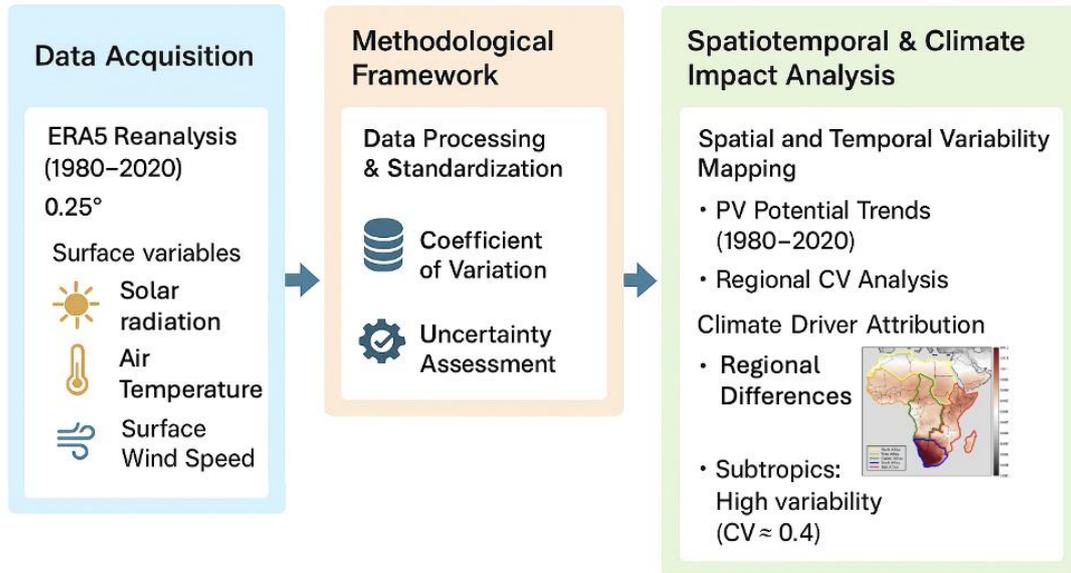

**Figure 1:** Methodological framework of the study. The workflow integrates ERA5 climate data extraction (1980–2020), empirical PV potential modeling, multivariate decomposition, and temporal stability assessment to quantify how radiation, temperature, and wind speed changes have individually and collectively influenced solar energy resources across Africa's diverse climate zones.

**2.2 Data and study area**

This study employs ERA5 reanalysis data (Hersbach et al., 2023), provided by the European Centre for Medium-Range Weather Forecasts through the Copernicus Climate Data Store. ERA5 offers global coverage at 0.25°×0.25° spatial resolution (~30 km at the equator) with hourly temporal resolution, making it one of the highest-resolution reanalysis products available for long-term climate studies. We analyze the period 1980–2020, providing four decades of continuous data suitable for assessing both long-term trends and interannual variability.

Three primary surface variables are extracted for PV potential calculations: (1) RSDS in W/m² representing total incoming solar irradiance; (2) TAS in °C governing module operating temperature and conversion efficiency; and (3) WS in m/s influencing convective cooling of PV panels. Spatial averaging is performed over five subregions—North Africa, West Africa, Central Africa, East Africa, and South Africa—defined by both climatic and geographical characteristics to capture regional heterogeneity.

ERA5 has been extensively validated against ground-based observations across Africa. For RSDS, typical biases range from −5% to +5% in most regions, with larger uncertainties (up to ±10%) in areas with complex terrain or persistent cloud cover (Sawadogo et al., 2020; Bloomfield et al., 2022). Temperature fields exhibit high accuracy (RMSE < 2°C), while wind speed shows greater variability (RMSE 1–2 m/s) due to the coarse representation of surface roughness (Neher et al., 2020). Despite these limitations, ERA5 represents the best available compromise between spatial coverage, temporal consistency, and physical realism for continent-scale solar resource assessment.

Our study area encompasses the African continent, spanning latitudes from 35°N to 35°S and longitudes from 20°W to 52°E. Africa exhibits exceptional climatic diversity, ranging from hyperarid deserts receiving over 280 W/m² mean annual solar radiation to humid equatorial forests with persistent cloud cover reducing radiation below 200 W/m². This spatial heterogeneity is driven by the continent's position straddling the equator, the influence of subtropical high-pressure systems, monsoon circulations, and complex topography.

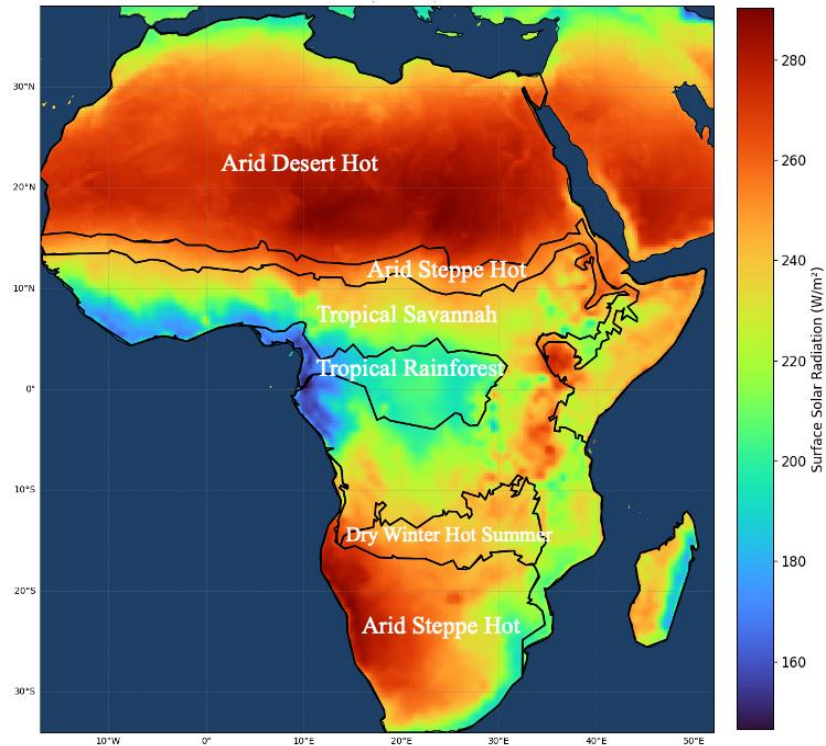

**Figure 2:** Solar radiation distribution and Africa climate zones. A gradient of RSDS is illustrated in unit of watts per square meter (W/m²). The major climate zones in Africa are labeled including the arid desert hot, arid steppe hot, tropical savannah, tropical rainforest and temperate dry winter hot summer regions.

Figure 2 illustrates the spatial distribution of mean annual solar radiation over 1980–2020 alongside Köppen-Geiger climate classifications, revealing the climatic foundation of Africa's solar resource potential. The highest radiation values exceeding 280 W/m² are concentrated in the Sahara Desert (hot desert) and the Kalahari-Namib region, corresponding to zones of persistent clear- sky conditions maintained by subtropical anticyclones. Intermediate values (220–260 W/m²) characterize semi-arid Sahel and East African savannas (Arid steppe hot), where seasonal cloud cover modulates annual totals. The lowest values below 200 W/m² occur in the Congo Basin (tropical rainforest climate), where convective cloud systems produce year-round attenuation of incoming solar radiation.

**2.3 PV Power Generation Potential**

PV potential (PVpot) is computed using an empirically-derived formula that accounts for the combined effects of radiation, temperature, and wind speed on conversion efficiency (Jerez et al., 2015):

$$\text{PV}_{\text{pot}} = \alpha_1 \cdot \text{RSDS} + \alpha_2 \cdot \text{RSDS}^2 + \alpha_3 \cdot \text{RSDS} \cdot \text{TAS} + \alpha_4 \cdot \text{RSDS} \cdot \text{WS} \quad (1)$$

where $\alpha_1 = 1.1035 \times 10^{-3}\,\text{m}^2/\text{W}$, $\alpha_2 = -1.4 \times 10^{-7}\,\text{m}^4/\text{W}^2$, $\alpha_3 = -4.715 \times 10^{-6}\,\text{m}^2/\text{W}°\text{C}$ and $\alpha_4 = 7.64 \times 10^{-6}\,\text{ms}/\text{W}$ in the corresponding units. These coefficients are derived from field measurements of monocrystalline silicon PV modules under diverse environmental conditions and have been validated across multiple climatic zones (Crook et al., 2011; Jerez et al., 2015).

**2.3.1 Climate Impact in PV Power Generation Potential**

To isolate the individual contributions of RSDS, TAS, and WS to observed PV potential changes, we employ the multivariate decomposition framework developed by Crook et al. (2011). For a change between two time periods (denoted by Δ), the total change in PVpot can be expressed as:

$$\Delta PV_{pot} = \Delta RSDS(\alpha_1 + \alpha_2 \Delta RSDS + \alpha_3 TAS + \alpha_4 VWS) + \alpha_3 RSDS \cdot \Delta TAS + \alpha_4 RSDS \cdot \Delta VWS + \alpha_3 \Delta RSDS \cdot \Delta TAS + \alpha_4 \Delta RSDS \cdot \Delta VWS \qquad (2)$$

This equation includes both first-order terms (representing direct effects of individual variables) and cross-product terms (capturing interactions between variables). The last two terms are typically small (<5% of total change) but are retained for completeness.

By setting RSDS and WS to zero in equation (2), the change in PVpot due to a change in TAS can be obtained. Analogously, the change in PVpot due to the influence of changes in WS alone is given by imposing RSDS and TAS to zero in equation (2). However, the cross-products in the last two terms of equation (2) introduce interactions between variables, complicating the isolation of individual contributions and necessitating the use of multivariate decomposition techniques. This decomposition is performed for each grid. Percentage contributions are calculated by dividing individual variable contributions by the total change in PVpot.

## 2.4 Statistical Analysis Methods

We employ various statistical methods to analyze changes in PV potential. We calculate the mean, maximum, and minimum values for PVpot, RSDS, TAS, and WS to characterize their spatial and temporal variations. The stability of PV potential in different regions is quantified using the CV, which is calculated as the ratio of the standard deviation to the mean value. CV has been previously employed in other studies to evaluate the variability of renewable energy, particularly to assess solar and wind resource stability. Additionally, we perform trend analysis using linear regression to identify significant changes over time.

## 3. Results

### 3.1 Spatial-Temporal Patterns of PV Potential Change

Figure 3 presents the spatial distribution of changes in PV potential and its underlying climate factors across Africa between 1980–1999 and 2000–2019. The continent exhibits a mean increase of 3.2% in PV potential (Figure 3a), but this aggregate masks pronounced spatial heterogeneity. East and Central Africa show the strongest gains, with increases exceeding 6% in localized areas of the Ethiopian Highlands and East African Rift Valley. In contrast, North Africa experiences slight declines of approximately 0.5%, particularly in the central Sahara, while South Africa displays mixed patterns with increases in coastal regions but declines in interior plateaus.

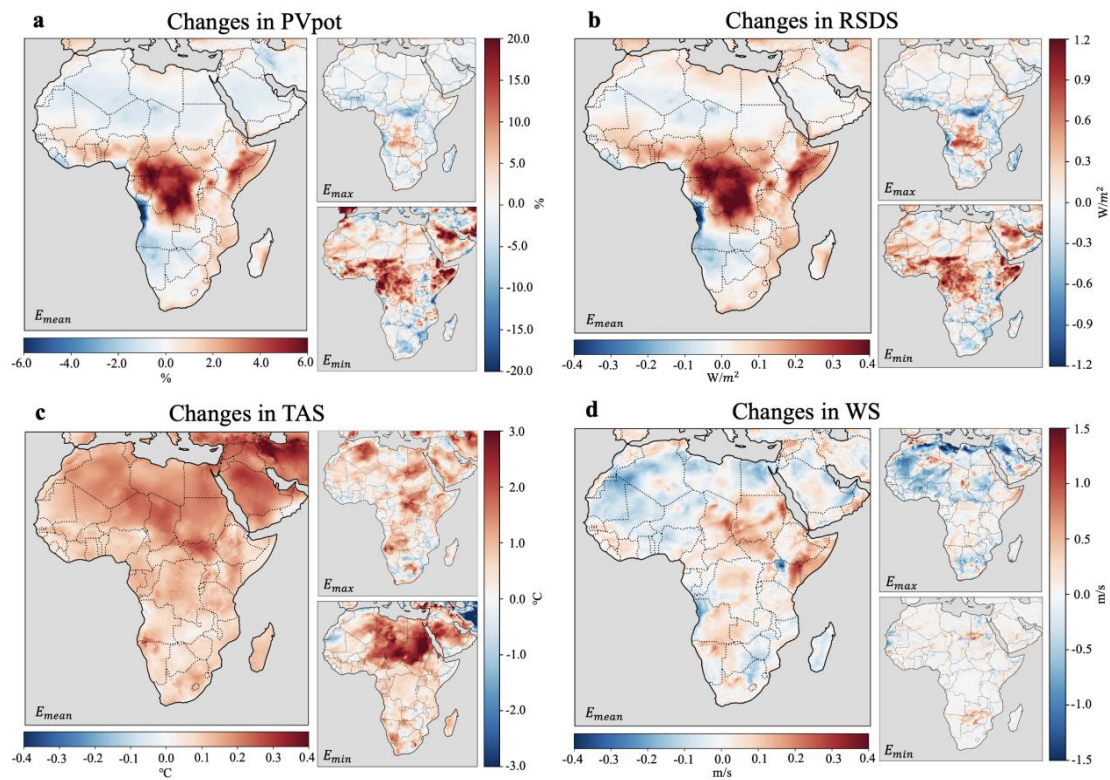

**Figure 3**: Decadal Changes in PV Potential and Climate Variables (1980–1999 vs. 2000–2019). The figure shows the spatial distribution of changes in (a) PVpot, (b) RSDS, (c) TAS, and (d) WS. Color scales represent percentage changes for PVpot, W/m² for RSDS, °C for TAS, and m/s for WS.

The spatial pattern of radiation change (Figure 3b) largely explains the PV potential distribution. East Africa shows widespread radiation enhancements of +0.4 to +0.6 W/m², directly driving the region's PV gains. North Africa exhibits modest radiation increases (+0.1 to +0.3 W/m²) that prove insufficient to offset thermal losses. Temperature increases are ubiquitous (Figure 3c), averaging +1.2 °C continent-wide, but with pronounced spatial variation. The most severe warming exceeds +2 °C in the Sahara and Kalahari deserts, corresponding to areas where PV potential declines despite radiation gains. Wind speed changes are modest and spatially heterogeneous (±0.1 m/s, Figure 3d), confirming this variable plays a secondary role in determining PV trends.

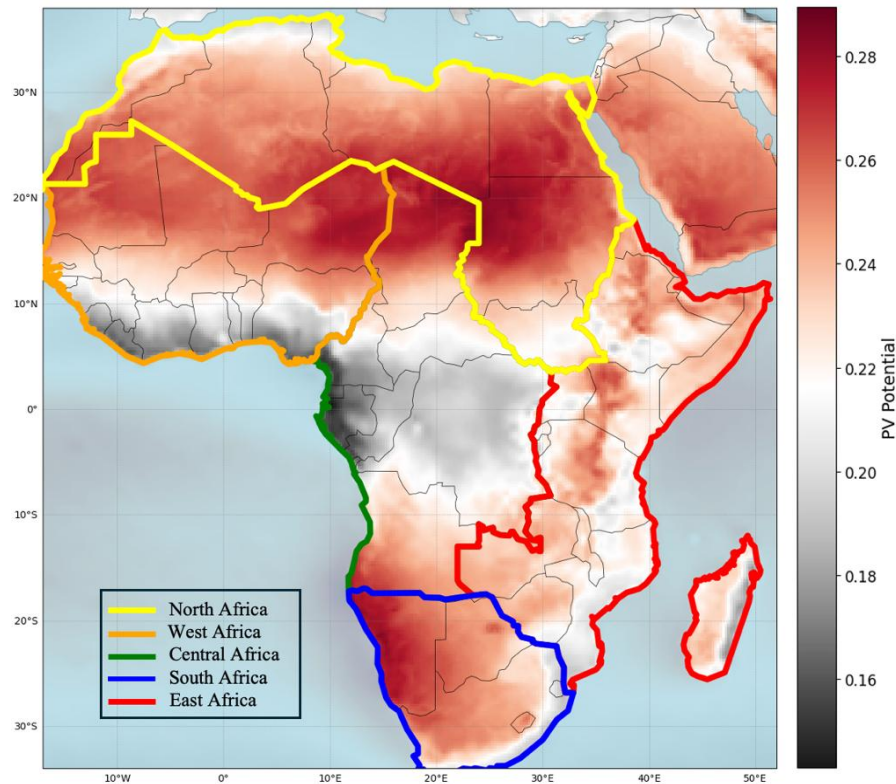

**Figure 4**: Regional PV Potential Distribution for 1981–2019. The figure shows the PV potential across African subregions (North Africa, West Africa, Central Africa, South Africa, East Africa).

The regional distribution of mean annual PV potential (Figure 4) reveals Africa's diverse solar resource base. North Africa dominates with the highest values, exceeding 0.25 throughout the region and reaching peaks above 0.28 in the central Sahara and Egyptian desert. South Africa's western regions exhibit comparably high values (>0.28) in the Kalahari and Northern Cape. East Africa shows moderately high values (0.23–0.25) in the Ethiopian Highlands, combining favorable irradiance with cooler elevated temperatures. Central

Africa presents the lowest potential (<0.225) due to persistent Congo Basin cloud cover, while West Africa displays a strong north-south gradient from high Sahel values (0.24–0.26) to lower coastal values (<0.22).

Temporal evolution over 1980–2020 (Figure 5) reveals contrasting regional trajectories. East Africa exhibits the most favorable trend, with PV potential rising steadily from approximately 0.220 to 0.235 (a relative increase of nearly 7%). This consistent upward trajectory, combined with low interannual variability, positions East Africa as the continent's most dynamically improving solar resource. North Africa demonstrates a concerning pattern: after maintaining values above 0.250 through the 1990s, the region's potential has stagnated since 2000, with recent years showing slight declines. This flattening indicates that rising temperatures are increasingly constraining the region's exceptional solar resource. South Africa shows the highest interannual variability, with values fluctuating between 0.245 and 0.275, reflecting sensitivity to ENSO and other climate teleconnections. Central and West Africa maintain stable values throughout the study period with no discernible long-term trends, suggesting that competing climate drivers roughly balance over multi-decadal timescales.

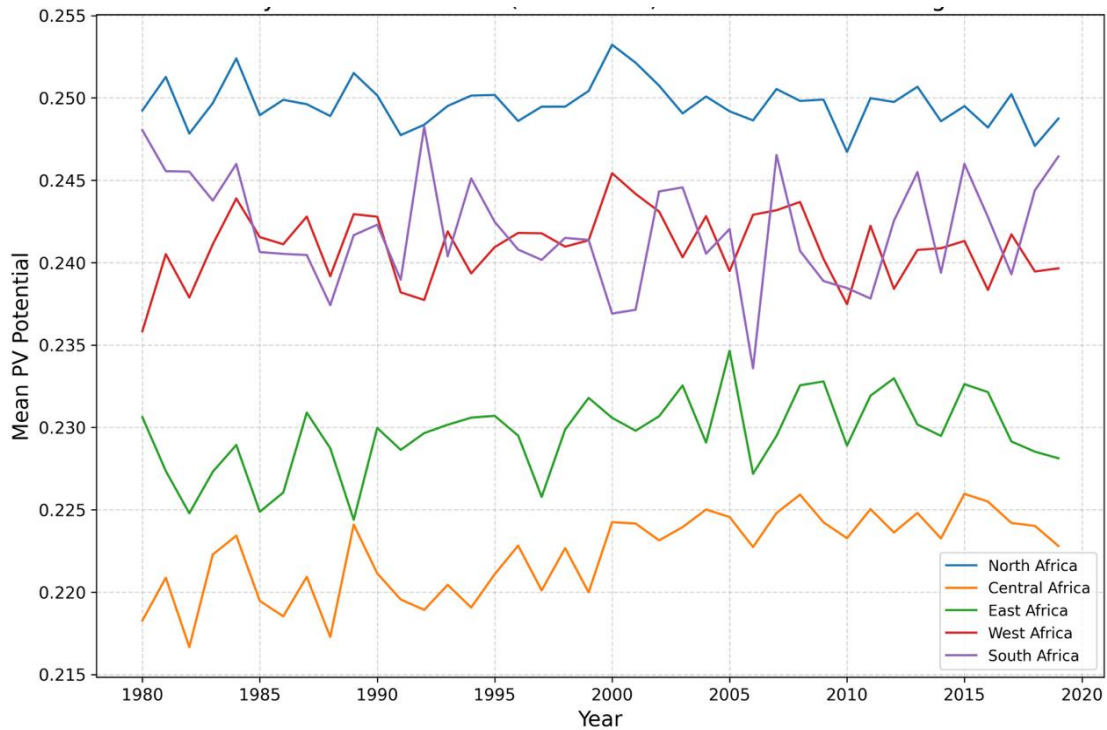

**Figure 5**. Temporal variations in PV potential across African subregions of North Africa, Central Africa, East Africa, West Africa, and South Africa from 1980 to 2020. This fig illustrates the regional differences and temporal changes in PV potential, providing insights for solar energy planning and development across the continent.

**3.2 Climate Driver Attribution and Seasonal Dynamics**

Multivariate decomposition quantifies the individual contributions of climate variables to PV potential changes (Figure 6). At the continental scale, radiation emerges as the dominant driver, contributing +72±8% of the observed gain, corresponding to a mean enhancement of +0.52 W/m² (Figure 6a). Temperature acts as the primary offsetting factor, contributing -23±5% through efficiency losses from +1.2°C mean warming (Figure 6b). Wind speed contributes modestly (+5±2%), providing slight cooling benefits in regions experiencing increases (Figure 6c).

Regional decomposition reveals pronounced spatial heterogeneity in driver contributions. In East Africa, radiation accounts for approximately +8% of PV potential change while temperature offsets only -1.5%, producing the net gain of +6%. This favorable balance reflects both strong radiation enhancement and moderate temperatures that buffer against warming impacts. North Africa presents a contrasting profile: radiation contributes +1% while temperature contributes -4%, yielding the net decline of -0.5%. This represents the continent's most unfavorable balance, where extreme temperatures amplify the negative efficiency impacts of additional warming. Central Africa shows minimal net change, with small positive radiation contributions (+0.5%) nearly balanced by small negative temperature contributions (-0.3%). South Africa exhibits high spatial variability in all components, generally showing radiation and temperature contributions of comparable magnitude that partially offset each other.

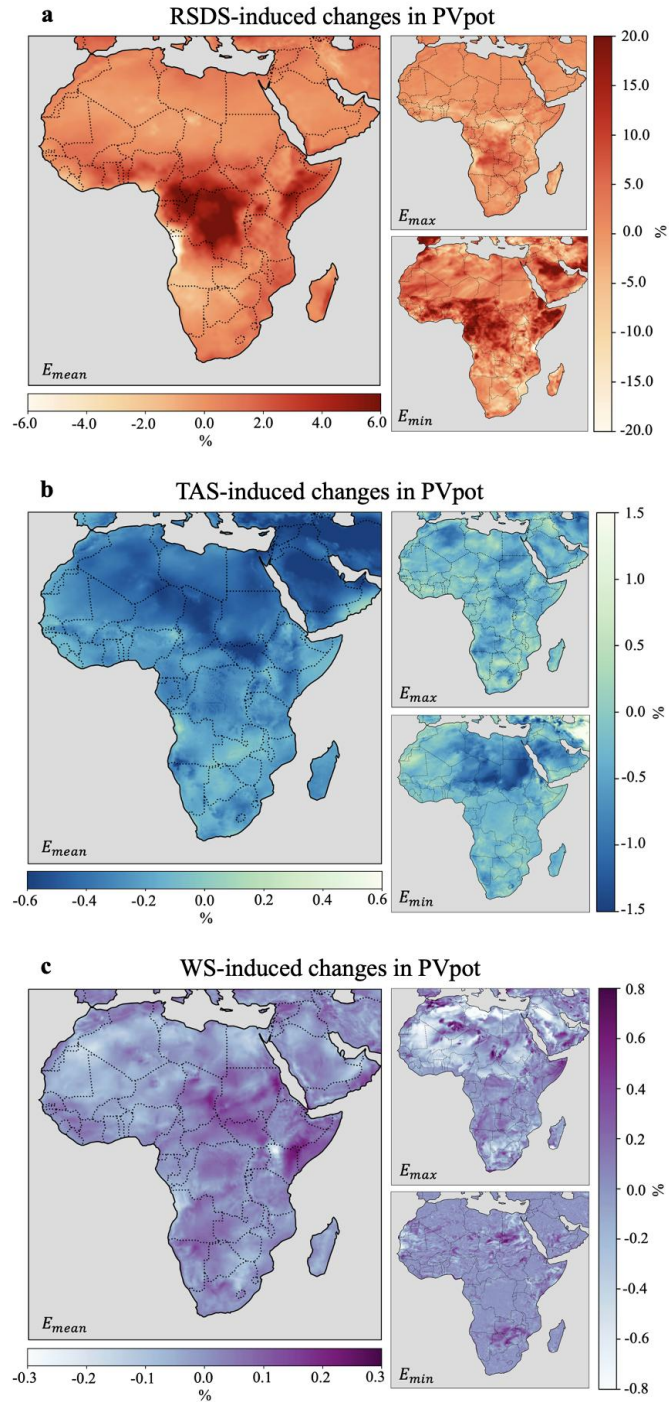

**Figure 6**: Impacts on PV potential induced by (a) RSDS (b) TAS (c) WS. Radiation dominates (+72±8%), followed by temperature (-23±5%) and wind (+5±2%). East Africa benefits from RSDS-driven gains (+8%), while North Africa suffers TAS-induced losses.

Seasonal variations (Figure 7) reveal pronounced intra-annual modulation of these climate drivers. PV potential anomalies (Figure 7a) show that East and Central Africa experience maximum positive anomalies (2–6%) during June–August, when clear skies and high solar elevation angles combine favorably. The Sahel region exhibits distinctive wet-season reductions (June–August), where monsoon-associated cloud cover suppresses PV output by approximately 5%. South Africa demonstrates consistent negative anomalies year-round (-0.2% to -0.4%), with peak losses during December–February summer when high temperatures severely degrade module efficiency.

Radiation anomalies (Figure 7b) drive much of the observed seasonal variation, with East Africa's summer peak reaching +0.6 W/m² and the Sahel's monsoon-season deficit exceeding -0.8 W/m². Temperature anomalies (Figure 7c) create consistent efficiency drags in subtropical zones, particularly in South Africa where year-round negative impacts intensify during summer. Wind speed anomalies (Figure 7d) show complex spatial patterns but limited magnitude, with East Africa exhibiting modest spring increases (+0.1–0.2 m/s) that provide beneficial cooling during transitional months.

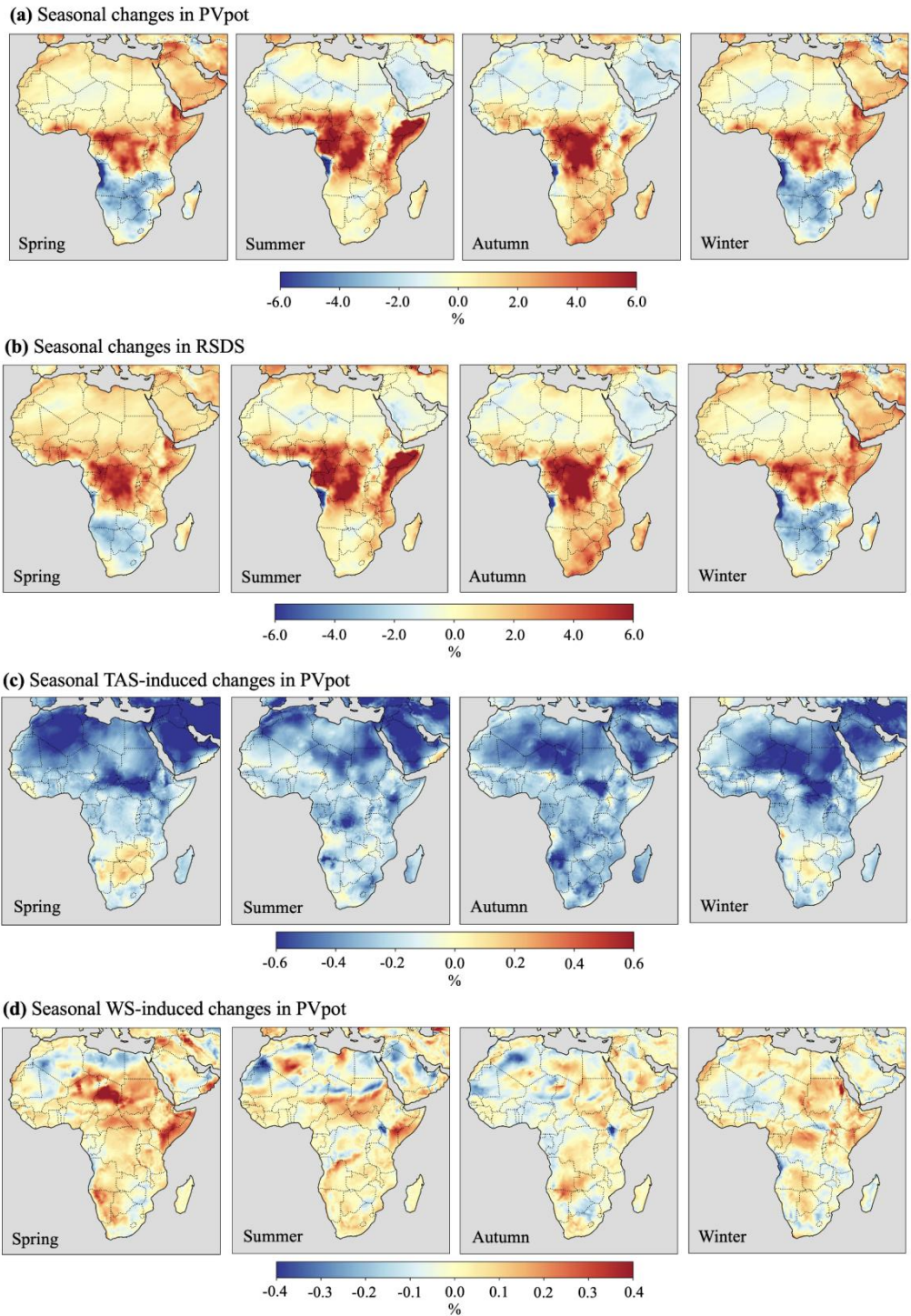

**Figure 7**: Seasonal PV Potential and Climate Variable Anomalies. Seasonal changes in (a) PVpot, (b) RSDS, (c) TAS-induced PVpot, and (d) WS-induced PVpot.

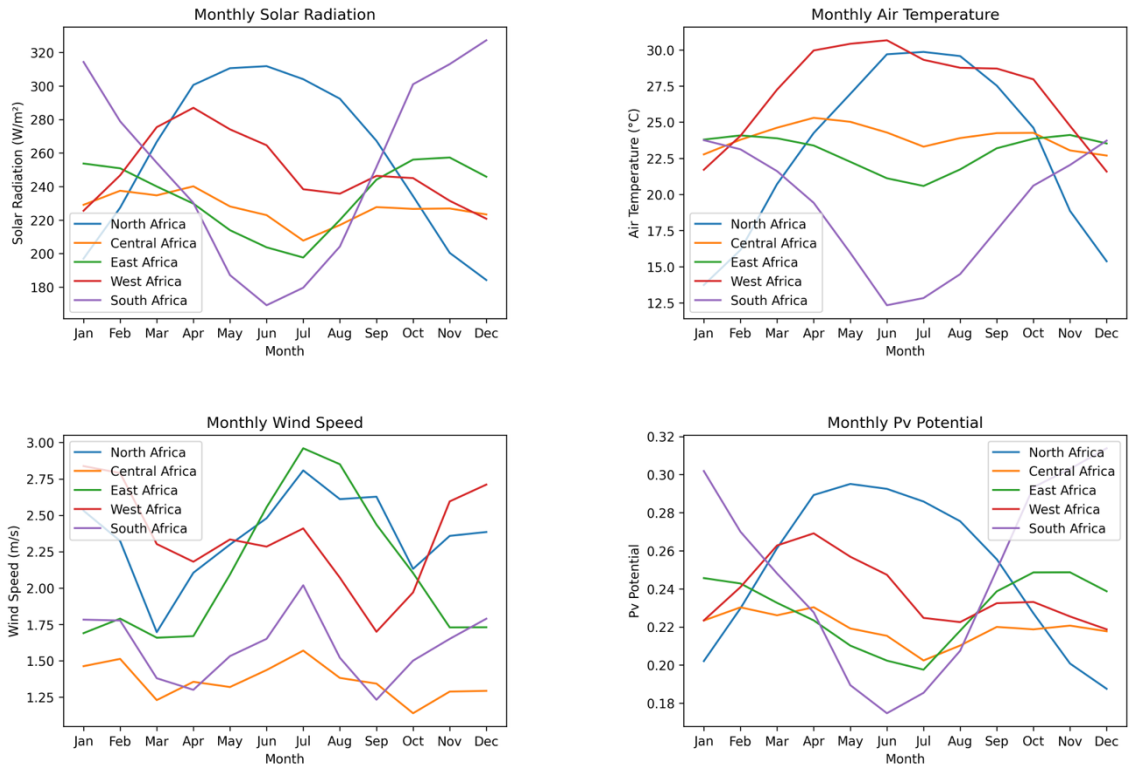

**Figure 8**: Monthly Variability of Climate Variables and PV Potential across African Regions of North, East, Central, South, and West Africa.

Monthly climatologies (Figure 8) provide detailed context for these seasonal patterns. North Africa maintains consistently high temperatures exceeding 25°C from April through October, peaking above 30°C in July–August, creating chronic thermal stress for PV systems. East Africa exhibits moderate and stable temperatures (17–25°C year-round) with minimal seasonal amplitude, maintaining optimal module operating conditions. South Africa displays the largest seasonal temperature range (12–24°C), with summer peaks imposing efficiency penalties but mild winters allowing near-optimal performance. Monthly radiation patterns confirm the dominant control of cloud cover regimes: Central Africa maintains consistently low values (200–220 W/m²) year-round, the Sahel shows pronounced monsoon-season suppression, and North Africa sustains the highest and most stable radiation (>280 W/m²) throughout the year. Wind speeds show modest monthly variation (3.5–4.5 m/s) across regions, confirming their secondary role in PV performance.

**3.3 Temporal Stability and Extreme Event Impacts**

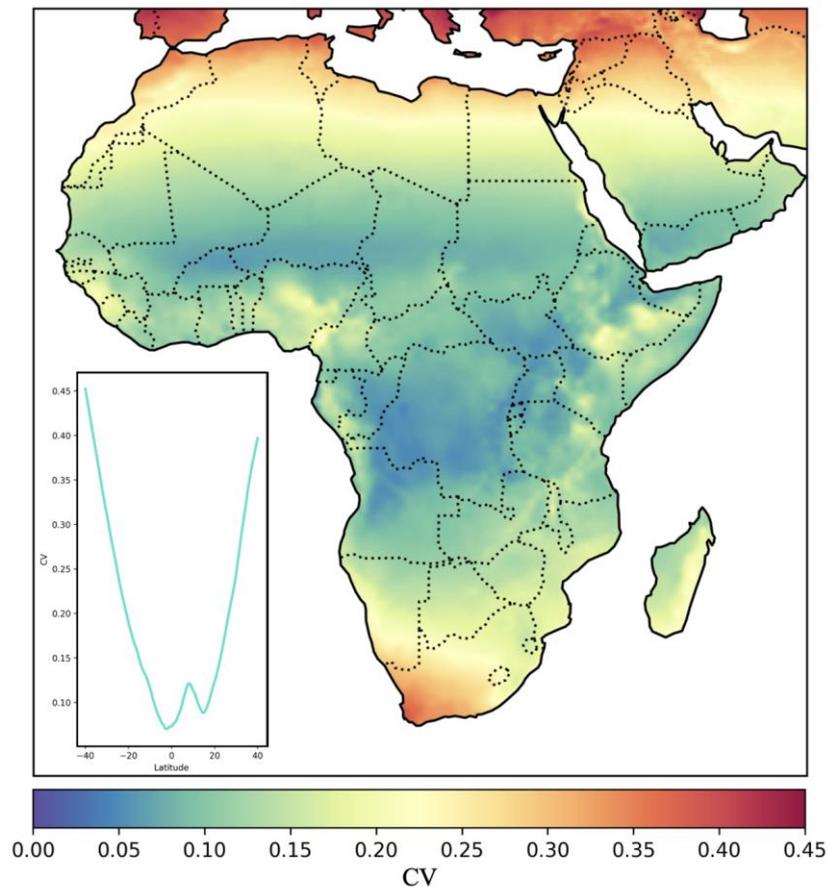

**Figure 9**: CV distribution of PV potential and its latitudinal Gradient. Equatorial zones (CV=0.1) show high stability, contrasting with subtropical regions (CV=0.4).

The CV analysis (Figure 9) reveals that high resource abundance does not guarantee operational stability. The continent exhibits a pronounced latitudinal gradient, with equatorial regions demonstrating remarkably low variability (CV≈0.1) and subtropical zones showing high variability (CV approaching 0.4). Central Africa and equatorial

East Africa form a stable core with CV values below 0.12, indicating year-to-year fluctuations typically less than 12% of the mean. This exceptional stability arises from consistent solar geometry and balanced atmospheric conditions, where compensating cloud cover effects create stable radiation regimes despite substantial precipitation.

North Africa displays CV values of 0.3 – 0.4, indicating substantial interannual variability despite possessing the continent's highest mean PV potential (>0.25, Figure 4). This instability reflects sensitivity to subtropical jet variability, dust storms, and heat extremes that modulate atmospheric transparency on interannual timescales. South Africa exhibits the continent's most variable solar resource, with CV values approaching 0.4 driven by sensitivity to ENSO, the Indian Ocean Dipole, and subtropical circulation shifts. West Africa shows intermediate variability (CV=0.2–0.3), higher in the Sahel due to monsoon and drought sensitivity, transitioning to lower coastal values with oceanic stabilization.

This stability gradient fundamentally challenges conventional resource assessment: North Africa's exceptional mean potential is substantially devalued by high variability, while East Africa's moderate potential (0.22–0.24) is enhanced by exceptional stability. For risk-averse investors, grid planners prioritizing reliability, or off-grid applications serving critical loads, stability may outweigh mean potential in determining optimal deployment locations.

Extreme climate events drive much of the observed variability (Figure 10). Heatwave frequency has increased significantly across all regions (Figure 10a), with East Africa showing the steepest trend (+0.082%/year) and North Africa close behind (+0.069%/year). Despite increasing frequency, individual heatwaves exert relatively modest direct impacts on PV potential (Figure 10b), reducing output by approximately 0.3% in North Africa during events. This occurs because heatwaves typically coincide with clear-sky conditions that enhance irradiance, partially offsetting temperature-induced efficiency losses.

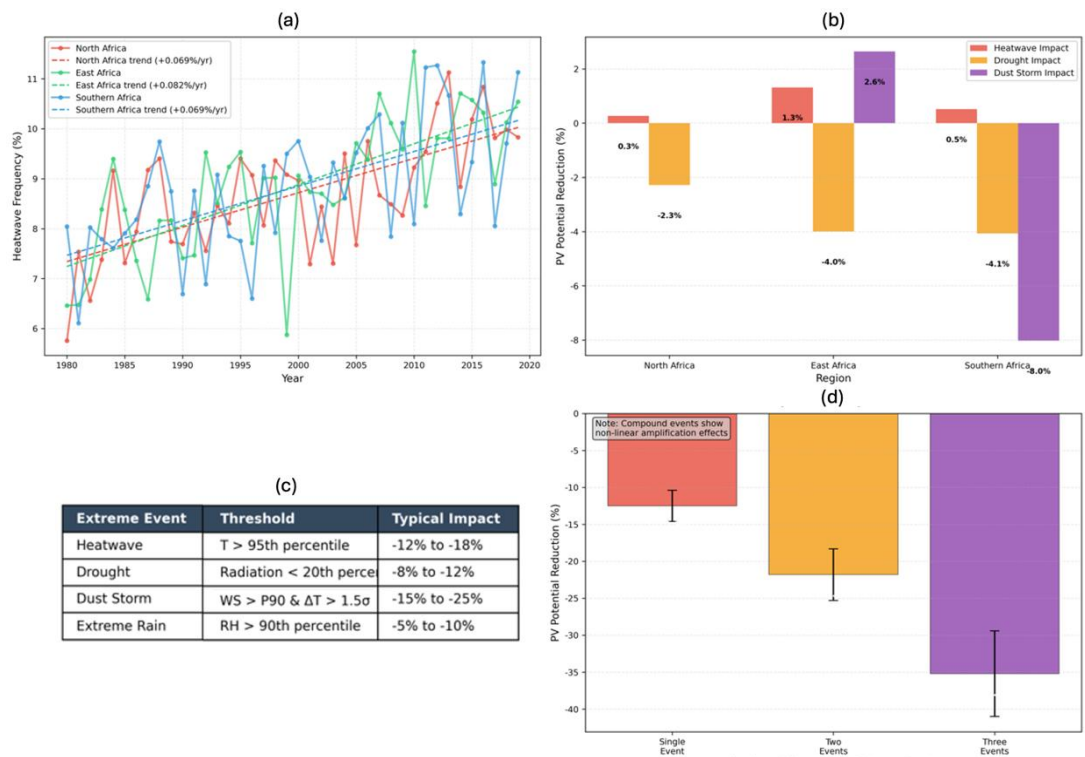

**Figure 10.** Impact of Extreme Climate Events on PV Potential in Africa. This figure investigates the influence of extreme climate events: (a) heatwave frequency trends, (b) PV potential reduction by extreme event type, (c) extreme climate event thresholds and impact, (d) cumulative impact of compound extreme events.

In contrast to heatwaves, drought and dust storm events impose more severe impacts (Figure 10c). Droughts reduce PV potential by 8–12% through increased dust loading and altered atmospheric conditions. Dust storms cause the most dramatic single-event impacts, reducing potential by 15–25% in the Sahel and Sahara through atmospheric attenuation and persistent panel soiling. Compound extremes produce the most severe impacts (Figure 10d). In South Africa, simultaneous heatwaves, droughts, and dust storms reduce PV potential by up to 40%, representing the most extreme climate-driven degradation observed continent-wide. These compound events are particularly challenging for grid management, creating sudden sustained generation deficits coinciding with elevated cooling loads.

The spatial distribution of extreme event impacts reveals that subtropical zones (North and South Africa) are most vulnerable, experiencing both the highest event frequencies and largest magnitude impacts. Equatorial regions show lower extreme event occurrence and smaller impacts, contributing to their superior stability profiles. The increasing frequency of extreme events suggests that stability may further deteriorate under continued climate change, underscoring the necessity of incorporating extreme event risk into solar project financing, insurance products, and grid resilience planning.

## 4. Discussion

### 4.1 Regional Divergence in Climate Impacts and Strategic Implications

Our analysis reveals a continental PV potential increase of 3.2% over four decades (1980–2020), but this aggregate conceals fundamental regional divergences that demand differentiated deployment strategies. Multivariate decomposition demonstrates that radiation enhancement accounts for 72±8% of the observed gain (corresponding to 0.52 W/m² mean increase), while temperature-induced efficiency losses offset 23±5% through 1.2°C mean warming. Wind speed contributes modestly (5±2%), primarily through convective cooling in regions experiencing increases. These findings confirm that solar irradiance remains the dominant driver of PV performance across Africa, but also reveal that temperature is transitioning from a secondary constraint to a primary limiting factor in already-hot arid zones.

The spatial heterogeneity of these impacts fundamentally challenges the conventional paradigm that equates high irradiance with superior solar potential. North Africa exemplifies this paradox: despite possessing the world's highest baseline irradiance exceeding 280 W/m² and mean PV potential above 0.25, the region experiences net declines of approximately 0.5% over the study period. Driver decomposition attributes this decline to severe thermal stress, where temperature contributes −4% to regional PV potential—nearly four times the positive contribution from radiation (approximately 1%). This unfavorable balance reflects extreme baseline temperatures exceeding 28°C that amplify the negative efficiency impacts of additional warming. Similar temperature-driven constraints have been documented in other hyperarid regions, including the Arabian Peninsula and Southwest Asia, where module operating temperatures frequently exceed 60°C during summer months, reducing conversion efficiency by 15–20% relative to standard test conditions (Hasan et al., 2015; Hudișteanu et al., 2024).

In stark contrast, East Africa emerges as the continent's most favorable zone for solar expansion, combining robust gains exceeding 6% in localized areas with exceptional temporal stability (CV approximately 0.1). Radiation enhancement contributes approximately 8% to regional PV potential while temperature offsets only 1.5%, yielding a net gain six times larger than North Africa's decline. This favorable balance arises from moderate baseline temperatures (17–25°C) that provide a critical thermal buffer: even with 0.8°C warming, module operating temperatures remain within optimal efficiency ranges. Additionally, East Africa's bimodal rainfall regime produces consistent year-round irradiance, avoiding the pronounced dry-wet contrasts that destabilize subtropical zones. These characteristics position East Africa as globally competitive with premier solar resources in the southwestern United States, Australia's interior, and Chile's Atacama Desert, but with the added advantages of lower temperature-induced losses and superior temporal stability.

Central and West Africa present intermediate scenarios. Central Africa exhibits the lowest mean PV potential (below 0.225) due to persistent Congo Basin cloud cover, but demonstrates remarkable temporal stability (CV approximately 0.1) indicating highly predictable output year-to-year. This stability arises from consistent equatorial solar geometry and balanced atmospheric conditions where cloud cover effects remain relatively constant despite substantial precipitation. West Africa displays strong north-south gradients, with high Sahel potential (0.24–0.26) moderated by pronounced seasonal variability during the June–September monsoon period when PV output decreases by approximately 5%. This seasonal bottleneck necessitates either energy storage, grid interconnection, or hybrid systems to maintain reliable supply during the wet season.

South Africa exhibits some of the continent's highest baseline potential (exceeding 0.28 in the Kalahari and Northern Cape), but also the most pronounced interannual variability (CV approaching 0.4). This instability is driven by sensitivity to ENSO teleconnections, the Indian Ocean Dipole, and subtropical circulation shifts that modulate cloud cover and atmospheric transparency on interannual timescales. Compound extremes—the concurrent occurrence of heatwaves, droughts, and dust storms—reduce PV potential by up to 40%, the most severe climate-driven degradation observed continent-wide. While South Africa's well-developed infrastructure provides opportunities for integrating climate-resilient technologies, the high variability demands sophisticated risk management including energy storage, over-sizing of generation capacity, and financial hedging instruments.

### 4.2 Strategic Framework for Climate-Resilient Solar Deployment

The spatial-temporal heterogeneities identified in this study demand a differentiated strategic framework that moves beyond simple capacity maximization to prioritize climate resilience and long-term reliability. We propose three regional archetypes with distinct deployment strategies:

High-Potential, High-Variability Zones (North and South Africa): These regions require heat-resilient technologies over capacity maximization. Priority interventions include advanced module cooling systems (passive and active), bifacial modules that leverage albedo from desert surfaces, tracking systems optimized for high-temperature operation, and dust mitigation strategies including automated cleaning systems and anti-soiling coatings. Grid-scale energy storage is essential to buffer interannual variability and manage compound extreme events. Regional grid interconnection can diversify supply by connecting high-potential but variable zones with more stable regions, reducing system-wide volatility. Financial instruments including weather derivatives and parametric insurance can hedge revenue risk from extreme events, making projects more bankable despite high variability.

Moderate-Potential, High-Stability Zones (East Africa and equatorial regions): These regions present opportunities for rapid, low-risk deployment at scale. The combination of favorable trends (East Africa's 7% increase over the study period), exceptional stability (CV approximately 0.1), and moderate temperatures creates conditions attractive to both development finance institutions and commercial investors. Priority actions include accelerated permitting and grid connection processes, blended finance mechanisms that de-risk early-stage projects to attract private capital, and regional cooperation frameworks that enable cross-border electricity trade. The reliability of these resources makes them particularly suitable for baseload-like generation that can displace fossil fuels and support industrial development.

Low-Potential, High-Stability Zones (Central and coastal West Africa): While these regions exhibit lower mean irradiance, their exceptional stability makes them well-suited for distributed generation and off-grid applications serving critical loads. Priority interventions include mini-grid and stand-alone systems for rural electrification, hybrid systems combining solar with complementary resources (hydro, biomass) to maintain year-round supply, and productive use applications (irrigation, cold storage, processing) that can operate flexibly around seasonal variations. The predictability of output in these regions reduces the storage requirements compared to high-variability zones, lowering overall system costs despite lower generation.

**4.3 Limitations and Future Research Directions**

Several limitations constrain the interpretation of our findings. While ERA5 reanalysis data are extensively validated, the 0.25° resolution may not fully capture fine-scale processes including aerosol loading, dust transport, orographic effects, and urban heat island effects that significantly influence local PV performance. Uncertainty analysis indicates that ERA5 RSDS exhibits typical biases of ±5% in tropical regions, translating to ±12% uncertainty in regional PV potential estimates. Sensitivity tests confirm that ±10% variation in RSDS produces ±8% change in PV potential, while ±1°C temperature variation yields ±2% change, indicating that radiation uncertainty dominates overall uncertainty.

Our PV performance model assumes constant technological parameters and does not account for evolving module materials, efficiency improvements, or degradation rates under different climate regimes. Next-generation technologies including perovskite cells, tandem architectures, and advanced thermal management systems may exhibit different climate sensitivities than the monocrystalline silicon modules parameterized in our model. Additionally, we do not consider socio-economic factors including grid infrastructure constraints, financing availability, policy environments, and land use conflicts that critically determine deployment feasibility beyond purely climatic considerations.

Future research should address these limitations through several pathways. Higher-resolution regional climate models (10–25 km) coupled with aerosol transport schemes can better resolve mesoscale processes and localized climate phenomena. Linking climate projections with techno-economic models would enable integrated assessments of deployment pathways that account for technology evolution, cost trajectories, and policy scenarios. Field validation campaigns measuring actual PV performance across diverse African climate zones would ground-truth model estimates and quantify operational losses from soiling, degradation, and grid curtailment not captured in climate-driven potential estimates. Finally, evaluating adaptation strategies including floating PV systems, agrivoltaics, and climate-responsive grid designs would provide actionable guidance for resilient infrastructure development.

**5. Conclusion**

This study provides a continent-wide assessment of climate change impacts on photovoltaic potential across Africa over four decades (1980–2020), revealing fundamental spatial-temporal heterogeneities that challenge conventional resource assessment paradigms and demand differentiated deployment strategies.

Our multivariate decomposition demonstrates that continental PV potential increased by 3.2%, driven primarily by enhanced solar radiation (0.52 W/m², accounting for 72±8% of the gain), but partially offset by warming-induced efficiency losses (1.2°C mean temperature increase, contributing −23±5%). However, this aggregate conceals profound regional divergences. East Africa emerges as the continent's solar frontier, combining robust gains exceeding 6% with exceptional temporal stability (CV approximately 0.1) and moderate baseline temperatures (17–25°C) that buffer against thermal stress. In stark contrast, North Africa—despite possessing the world's highest baseline irradiance exceeding 280 W/m² and mean PV potential above 0.25—experiences net declines of approximately 0.5% as extreme warming exceeding 2°C overwhelms radiation benefits. This paradox fundamentally challenges the assumption that resource abundance ensures superior deployment potential.

The temporal stability analysis reveals a critical dimension missing from conventional assessments. Equatorial regions demonstrate CV values below 0.12, indicating highly predictable year-to-year output with fluctuations typically less than 12% of the mean. Subtropical zones exhibit CV values of 0.3–0.4, indicating four-fold higher variability driven by sensitivity to extreme events and climate teleconnections. Compound extremes—the concurrent occurrence of heatwaves, droughts, and dust storms—reduce PV potential by up to 40% in South Africa, representing the most severe climate-driven degradation observed continent-wide. These findings establish that stability must be weighted alongside mean potential in resource valuation, particularly for risk-averse investors and off-grid applications serving critical loads.

These spatial-temporal heterogeneities demand a strategic reframing of Africa's solar deployment pathways. For North Africa and other high-potential but thermally constrained zones, the priority must shift from capacity maximization to heat-resilient technology deployment, including advanced cooling systems, bifacial modules, and dust mitigation strategies. Subtropical zones with high variability require grid-storage co-investment and regional interconnections to buffer interannual fluctuations and manage compound extreme events. East Africa and equatorial regions present opportunities for rapid, low-risk deployment at scale, leveraging their favorable trends, exceptional stability, and moderate temperatures to anchor regional power pools and industrial development.

The increasing frequency of extreme events across all categories—heatwaves rising at 0.069–0.082% per year, dust storms intensifying in the Sahel, and compound extremes becoming more severe—suggests that climate impacts on solar resources will intensify in coming decades. Incorporating climate resilience into deployment strategies today is not merely prudent risk management but essential to ensuring that infrastructure built in the 2020s and 2030s remains productive through mid-century and beyond. This necessitates integrating extreme event risk into project financing, developing parametric insurance products, and designing grid systems with sufficient flexibility to manage sudden sustained generation deficits.

By quantifying climate driver contributions and resolving spatial-temporal heterogeneities at continental scale, this analysis provides an actionable framework for climate-resilient solar deployment. The findings demonstrate that Africa possesses sufficient stable, high-quality solar resources to support ambitious electrification goals—addressing energy access for 600 million people without electricity while pursuing low-carbon development pathways. However, realizing this potential requires moving beyond uniform expansion strategies to embrace regional differentiation that accounts for climate impacts, temporal stability, and evolving extreme event risks.

Future research should advance this framework through three priority pathways. First, higher-resolution regional climate models (10–25 km) coupled with aerosol transport schemes can resolve mesoscale processes and localized phenomena inadequately captured in reanalysis data. Second, linking climate projections with techno-economic models would enable integrated assessments of deployment pathways that account for technology evolution, cost trajectories, and policy scenarios. Third, field validation campaigns measuring actual PV performance across diverse African climate zones would ground-truth model estimates and quantify operational losses from soiling, degradation, and grid curtailment not captured in climate-driven potential assessments.

Ultimately, this study demonstrates that climate change is not merely an environmental context for Africa's energy transition—it is a fundamental determinant of where, how, and how reliably solar resources can be deployed. By providing quantitative evidence of regional climate impacts and stability patterns, our analysis equips policymakers, investors, and grid planners with the knowledge necessary to design climate-resilient infrastructure that can accelerate Africa's energy transition while building adaptive capacity to manage future climate variability. The continent's solar potential remains vast, but unlocking it sustainably requires strategies as diverse as Africa's climates themselves.

**CRediT authorship contribution statement**

Eva Lu: Conceptualization, Writing—original draft, Methodology, Formal analysis, Funding acquisition, Project administration. Dongdong Wang: Resources, Administration, Writing—review.

**Declaration of Competing Interest**

The authors declare that they have no known competing financial interests or personal relationships that could have appeared to influence the work reported in this paper.

**Data availability**

The ERA5 hourly datasets are available via the websites of the dataset producers (https://cds.climate.copernicus.eu/).